
\input phyzzx
\overfullrule=0pt

\def\p{\partial}
\def\ge{\varepsilon}

\rightline {UTTG-27-92}
\rightline {November 1992}
\bigskip\bigskip
\title{Black hole formation in $c=1$ String Field Theory
\foot{Work supported in part by NSF grant PHY 9009850 and
R.~A.~Welch Foundation.}}

\vskip 3cm
\author{Jorge G. Russo,\foot{e-mail: russo@utaphy.ph.utexas.edu } }
\address {Theory Group, Department of Physics, University of
Texas\break
Austin, TX 78712}
\vskip 2cm

\abstract
\singlespace
A suggestion on how black holes may appear in
Das-Jevicki Collective field theory is given.
We study the behaviour of a `test' particle when energy is
sent into the system.
A perturbation moving near the potential barrier can
create a large-distance black hole geometry where the seeming
curvature singularity is at the position of the barrier.
In the simplest `static' case the exact $D=2$ black hole
metric emerges.

\vfill\endpage

\Ref\mm {D. Gross and N. Milkovic, Phys. Lett. 238B (1990) 217; E.
Brezin, V. Kazakov and Al. B. Zamolodchikov, Nucl. Phys. B338 (1990) 673;
G. Parisi, Phys. Lett. 238B (1990) 209; P. Ginsparg and J. Zinn-Justin,
Phys. Lett. 240B (1990) 333.}

\Ref\dj {S.R. Das and A. Jevicki, Mod. Phys. Lett. A5 (1990) 1639.}

\Ref\POLCO{J. Polchinski, Nucl. Phys. B346 (1990) 253}

\Ref\gk{D. Gross and I. Klebanov, Nucl. Phys. B352 (1990) 671;
A.M. Sengupta and S.R. Wadia, Int. J. Mod. Phys. A6
(1991) 1961.}

\Ref\GKLEB{D. Gross and I. Klebanov, Nucl. Phys. B359 (1991) 3.}

\Ref\GKN{D. Gross, I, Klebanov and M. Newmann, Nucl. Phys. B350 (1991)
621; U. Danielsson and D. Gross, Princeton Preprint PUPT-1258 (1991).}

\Ref\pol {J. Polchinski, Nucl. Phys. B362 (1991) 125.}

\Ref\djr{K. Demeterfi, A. Jevicki and J.P. Rodrigues, Nucl. Phys. B362
(1991) 173 and Nucl. Phys. B365 (1991) 199;
G. Mandal, A.M. Sengupta and S.R. Wadia, Mod. Phys. Lett. A6
(1991) 1465;
G. Moore, Nucl. Phys. B368 (1992) 557; G. Moore, M.R. Plesser
and S. Ramgoolam, Nucl. Phys. B377 (1992) 143.}

\Ref\AJEV{J. Avan and A. Jevicki, Phys. Lett B266 (1991) 35; Phys. Lett.
B272 (1990) 17; Mod. Phys. Lett. A7 (1992) 357.}

\Ref\DDMW{S.R. Das, A. Dhar, G. Mandal and S.R. Wadia, Int. J. Mod. Phys.
A7 (1992); Mod. Phys. Lett. A7 (1992) 71; Mod. Phys. Lett. A7 (1992) 937.}

\Ref\MORSEIB{G. Moore and N. Seiberg, Int. J. Mod. Phys. A7 (1992) 2601.}

\Ref\EGUCHI{ T. Eguchi, H. Kanno and S.K. Yang, Newton Institute Preprint
NT-92004 (1992).}

\Ref\tsey{A.A. Tseytlin, Int. J. Mod. Phys. A5 (1990) 1833.}

\Ref\wittbh{E. Witten, Phys. Rev. D44 (1991) 314.}

\Ref\maswa{G. Mandal, A. Sengupta and S.R. Wadia, Mod. Phys. Lett. A6
(1991) 1685}

\Ref\DVV{R. Dijkgraaf, E. Verlinde and H. Verlinde, Nucl. Phys. B371
(1992) 269. }

\Ref\brusell{J. Ellis, N.E. Mavromatos and D. Nanopoulos, Phys. Lett. B272
(1991) 261; Phys. Lett. B267 (1991) 465;
R. Brustein and S. de Alwis, Phys. Lett. B272 (1991) 285.}

\REF\witten{E.~Witten, IAS preprint, IASSNS-HEP-92/24.}

\REF\dasdahr{S. R. Das, Tata Institute preprint, TIFR-TH-92/62 (1992);
A. Dhar, G. Mandal and S.R. Wadia, Tata Institute preprint, TIFR-TH-92/63
(1992). }

\Ref\DFKUT{P. Di Francesco and D. Kutasov, Phys. Lett. B261 (1991) 385.}

\Ref\POLY{A. Polyakov, Mod. Phys. Lett. A6 (1991) 635.}

\Ref\MART{E. Martinec and S. Shatashvili, Nucl. Phys. B368 (1992) 338.}

\Ref\Wit{E. Witten, Nucl. Phys. B373 (1992) 187; I. Klebanov and
A. Polyakov, Mod. Phys.Lett. A6 (1991) 3273.}

\Ref\ark{A.A. Tseytlin, Phys. Lett. B268 (1991) 175.}

The double scaling continuum limit of the $c=1$ matrix model
[\mm-\EGUCHI ]
is expected to describe critical strings moving in two space-time
dimensions (see e.g. refs. [\POLCO, \tsey]).
Since the discovery of a black hole background
in this continuum two-dimensional theory [\wittbh -\DVV]
there have been a number
of attempts to find the counterpart in the matrix model formulation
[\brusell -\dasdahr ]\foot{The present approach
is different from other interesting, alternative proposals
(see refs.[\brusell ,\dasdahr ]). }.

In the last two years there was considerable progress in understanding
another $D=2$ string background,
the Liouville theory coupled to one scalar matter field [\pol,
\MORSEIB, \DFKUT -\Wit]. In particular, this continuum theory was shown to
have the same $w_\infty $ symmetry that appears in the $c=1$
matrix model
[\AJEV -\EGUCHI, \Wit ], and a good agreement between the tree-level
S-matrices was found (see e.g. refs. [\DFKUT , \POLY ]).

The matrix model is equivalent to a $(1+1)$-dimensional field theory
where the collective field is the density of eigenvalues of the matrix
[\dj ]. This theory is expected to be the field theory of the only
propagating mode of $D=2$ string theory, where there are no
transversal excitations. At classical level, one expects a relation
with the string effective action after all other modes of the spectrum
have been gauged away or integrated out by the equations of motion.
The absence of a metric in a resulting field theory
obscures the geometrical interpretation of physical processes.

To get a rough idea of what are the signs which would
indicate the presence of a black hole in the DJ field theory,
it is instructive to perform the integration of metric and dilaton in the
$D=2$ string theory, even if this will be done only in a very approximative
way.

To leading $\alpha '$ order, the tree-level string effective action for
the metric, dilaton and tachyon is given by,
$$
S = \int d^2x \sqrt{-G}
\bigl[e^{-2\phi}(R+4\p_\mu\phi\p^\mu \phi+c)
-\p_\mu \eta\p^\mu \eta -(\p_\mu\phi\p^\mu\phi-\nabla^2\phi
+{2\over \alpha '})\eta ^2+O(\eta ^3)] \ ,
\eqn\eff
$$
where $\eta$ is related to the usual tachyon
by $\eta =\exp (-\phi)T$, and $c=-8/\alpha' $.

In the conformal gauge $G_{\mu\nu}=e^{2\rho }\eta_{\mu\nu}$
($\eta_{00}=-\eta_{11}=1$),
the equations of motion take the form
$$
\p_+\p_- (\rho -\phi)= -{1\over 2}\p_+(e^\phi\eta )\p_-(e^\phi\eta )
\ \ , \eqn\equ $$
$$4\p_+\p_- e^{-2\phi} = c e^{2\rho -2\phi} - {2\over \alpha '}
e^{2\rho }\eta^2
\ \ , \eqn\eqd $$
$$\p^2_\pm \phi -2\p_\pm \rho \p_\pm\phi= {1\over 2}\p_\pm(e^\phi\eta)
\p_\pm (e^\phi\eta)\ \ ,
\eqn\eqt $$
$$
\p_+\p_-\eta= (\p_+\phi\p_-\phi-\p_+\p_-\phi+{1\over 2\alpha'}
e^{2\rho})\eta^2 +O(\eta^3) \ \ . \eqn\tachy
$$
Ignoring the $\eta $ back reaction the general solution to eqs.\equ -\eqt
\ is the black hole background discussed in refs.[\wittbh - \DVV ].
Inserting this background into eq. \tachy  \ one can see that $\eta $ becomes
massless far away.
By making perturbation theory around the linear dilaton background,
the $\eta $ dependence in eqs.\equ, \ \eqd \  can be neglected
in the first order approximation.
So let us assume that far away we have $\p_+\p_-(\rho-\phi )\sim 0$
and hence we can fix the ``Kruskal" gauge $\rho \sim \phi $.
Then eq. \eqd\   becomes, $4\p_+\p_- e^{-2\phi }
\sim c $, i.e. $e^{-2\phi }={c\over 4}x^+x^- +h_+(x^+)+h_-(x^-) $.
The leading piece of the constraint equations \eqt \ takes the form
(we assume $\p_\pm\eta >>\eta \p_\pm\phi\sim -\eta/2x^\pm $
for $x^\pm\to \pm \infty $)

$$\p_\pm ^2 e^{-2\phi } \sim -\p_\pm \eta \p_\pm\eta\ .$$
Therefore $h_\pm '' \sim - \p_\pm \eta\p_\pm \eta $, or
$$h_\pm (x^\pm)\sim - \int \int \p_\pm\eta\p_\pm\eta \ \ .
\eqn\constr
$$
Inserting these solutions for $\rho $ and $\phi $ into
eq. \tachy  , what remains is a nonlinear integro-differential equation
expressed purely in terms of $\eta $. From this equation
one can obtain scattering amplitudes corresponding to this
scalar field effective theory.

To see gravitational collapse we imagine that an energetic
 wave  $\eta_0 $, $\p_+\p_-\eta_0 \sim 0 $ is sent in, expand
$\eta =\eta_0 +\ge$, and study the behaviour of the small
fluctuation or `test' particle $\ge $
in the incoming background $\eta_0$. Then the conformal factor
and dilaton become $e^{-2\rho}=e^{-2\phi}={c\over 4}x^+x^-
+h_+(\p_+\eta_0) + h_-(\p_-\eta_0) +O(\ge )$\ , which
for a large class of $\eta_0$ represents a black hole.
Then eq. \tachy\  takes  the form
$$\p_+\p_- \ge = m(\eta_0)\ge +j(\eta_0) +O(\ge^2)\ \ ,
\eqn\fluc
$$
where both $m, \ j\to 0 $ asymptotically. The black hole geometry has to be
read out of $m(\eta _0) $ in eq.\fluc , since in the conformal gauge
it does not show up in the Laplacian.
However, by repeating the above procedure, e.g., in
the linear dilaton gauge $\phi=-x$
one finds a Laplacian of the form $f^{-1}\p^2_t \ge - f\p^2_x\ge +O(\ge ^2)$
,\ \ $f\equiv 1-M(\eta_0 )e^{-2x}\ , $
where the underlying large-distance geometry is exhibited in a manifest way.
\bigskip

The bosonic hamiltonian for the double scaled $c=1$ matrix model
is given by [\dj ] (we follow the notation of ref.[\pol ] )
$$
H=\int dx \bigg( {g^2_s\over 2}\Pi_\zeta (\p _x\zeta )\Pi_\zeta
+{\pi^2\over 6g^2_s }(\p _x\zeta )^3 +{1\over 2g_s^2}
(1-x^2)(\p _x\zeta )\bigg)\ \ ,
\eqn\ham
$$
where $x$ denotes the (rescaled)
space of eigenvalues of the original matrix model,
$\Pi_\zeta (x)$ is the momentum conjugate to $\zeta (x)$.
The equations of motion which follow from the above hamiltonian
are
$$
\p_t \zeta =g^2_s \Pi_\zeta\p_x\zeta \ ,\ \ \
\p_t\Pi_\zeta = {g^2_s\over 2}\p_x\Pi^2_\zeta +
{\pi^2\over 2g^2_s}\p_x(\p_x \zeta )^2 -{1\over g^2_s} x\ \ .
\eqn\hequs
$$
The general solution to these equations is [\pol]
$$
\Pi_\zeta =-{p_+ +p_-\over 2g^2_s}\ \ ,\ \ \
\p_x\zeta ={p_+ - p_-\over 2\pi } \ \ ,
\eqn\sol
$$
where $p_\pm (x,t)=a(\sigma_\pm ) \sinh (t-\sigma_\pm)$, $a(\sigma )$
is an arbitrary function of $\sigma $
and $\sigma_\pm=\sigma _\pm (x,t)$ are
the two solutions of $x=a(\sigma )\cosh (t-\sigma )$.

The momenta $p_\pm $ obey the transport equation
$$
\p_t p_\pm =x-p_\pm \p_x p_\pm \ \ \ .
\eqn\trans
$$
The ground state is given by the static solution
$$
\Pi_\zeta=0\ \ , \ \ \ \p_x\zeta _0={1\over\pi}(x^2-1)^{1/2}\ ,\ \
|x|>1\ \ \ \ .
\eqn\stat
$$
Now we introduce the scalar field $\Psi $ and its momentum
conjugate $\Pi $ as
$$
\zeta=\zeta _0 (x) + {g_s\over \sqrt{\pi}} \Psi \ \ ,\ \ \
\Pi_\zeta={\sqrt\pi\over g_s} \Pi \ \ \ .
\eqn\fluc
$$
After introducing a new coordinate $q$ as $x=\cosh(q)$
(we consider the right hand side of the barrier and $0\leq q <\infty$)
the hamiltonian \ham \ takes the form
$$
H={1\over 2} \int dq \bigg( \Pi ^2 + (\p_q \Psi )^2 +
{g_s\sqrt{\pi }\over \sinh ^2(q)}\big( \Pi^2\p_q \Psi +{1\over 3}
(\p_q \Psi )^3 \big)\bigg) \ \ \ .
\eqn\hamil
$$
The equations of motion for $\Psi $ and $\Pi $ are
($[\Psi (q,t),\Pi (q',t)]=i\delta (q-q') $ )
$$
\p_t \Pi=\p^2_q \Psi + {g\over 2}\p_q {\Pi ^2 +(\p_q \Psi
)^2\over \sinh^2 (q)}
\ \ ,\ \ \
\p_t \Psi = \Pi (1+  {g \over \sinh^2 (q)} \p_q \Psi  ) \ \ ,
\eqn\pies
$$
where $g\equiv g_s\sqrt{\pi} $. Eliminating $\Pi $ from these equations
one easily obtains
$${1\over A(\Psi )}\p_t^2 \Psi - A(\Psi ) \p_q^2 \Psi  =F(\Psi ) \ \ ,$$
$$F(\Psi )\equiv {2g(q)\over A(\Psi )^2}\p_t \Psi \p_q\p_t \Psi
-{g(q)\over A(\Psi )^3}(\p_t \Psi )^2 \p_q(g(q)\p_q \Psi )
-{g(q)\over \tanh (q)}({1\over A(\Psi )^2} (\p_t \Psi )^2 +(\p_q \Psi )^2)
\ , \eqn\seq
$$
where
$$
A(\Psi )\equiv 1+g(q)\p_q \Psi \ ,\ \ g(q)\equiv {g\over \sinh^2(q)} \ \ .
\eqn\aaa
$$
Here we do not assume that $\Psi $ is small and thus we shall keep
all higher powers in $\Psi $ in equation \seq .

For large $q$ the equation simplifies to
$$
(\p_t^2 -\p_q^2)\Psi  \sim 0 \ \ ,
\eqn\free
$$
with solution $\Psi=\Psi_+(t+q) + \Psi_- (t-q)$.

Along similar lines as in the first part of this note, now
we assume a physical situation in which there is an incoming
wave $\Psi_0$ and study the dynamics of  small fluctuations,
$$
\Psi=\Psi_0 + \ge\ \ \ ,\ \ \  \ \ge <<\Psi_0 \ .
\eqn\expa
$$
We shall be interested only in large distance physics,
where one expects to find analogous
results as those coming from the $\alpha '$ expansion
of the $D=2$ continuum string theory.

Inserting the expansion \expa \ in \seq \ and retaining only
the linear terms in $\ge $ we find
$$
{1\over A(\Psi _0)}\p_t^2 \ge - A(\Psi_0)\big( 1-
({g(q)\p_t\Psi_0\over A^2(\Psi_0)})^2\big) \p_q^2 \ge -
{2g(q)\over A(\Psi_0)^2}\p_t \Psi_0\p_q\p_t \ge +
Q\p_q \ge +T\p_t \ge \ $$
$$=A(\Psi_0)\p_q^2\Psi_0 - A^{-1}(\Psi_0)\p_t^2\Psi_0 +F(\Psi_0 )\ ,
 \eqn\eqge $$
where
$$Q\equiv -\p_q(g(q)\p_q \Psi_0 )-{g(q)\over A^2(\Psi_0)}\p_t^2 \Psi_0
+\p_q \big( {g^2(q)\over A^3(\Psi_0)}\big)(\p_t \Psi_0)^2 +
{4g^2(q)\over A^3(\Psi_0)}\p_t\Psi_0\p_q\p_t\Psi_0 \ , $$
$$T\equiv -\p_q\big( {g(q)\over A^2(\Psi_0)}\big)
\p_t\Psi_0- {2g(q)\over A^2(\Psi_0)}\p_q\p_t \Psi_0  \ \ .
$$

It is remarkable that the $\p_q\ge $ and $\p_t\ge $ terms in
eq. \eqge \ are such that this equation can be written in the form
$$
G_{\rm eff}^{\mu\nu}\nabla_\mu\p_\nu \ge = J_{\rm eff} +O(\ge ^2) \ \ ,
\eqn\effe
$$
where
$$
G_{\rm eff}^{\mu\nu}=\left(\matrix{A^{-1}(\Psi_0)& -
{g(q)\over A^2(\Psi_0 )}\p_t\Psi_0  \cr
 -{g(q)\over A^2(\Psi_0 )}\p_t\Psi_0 &
- A(\Psi_0)\big( 1-({g(q)\p_t\Psi_0\over A^2(\Psi_0)})^2\big) \cr}\right)
\ ,\ \
\eqn\metrica
$$
$$
J_{\rm eff}=A(\Psi_0)\p_q^2\Psi_0 - A^{-1}(\Psi_0)\p_t^2\Psi_0 +F(\Psi_0 )
\ \ .
\eqn\corr
$$
An effective, large-distance geometry has emerged.
The geometrical interpretation may break
down in the vicinity of the wall
($q=0$) where the $O(\ge ^2)$ terms can no longer be ignored.
A curious fact is that in these coordinates we have
$\det G_{\rm eff} =-1$ automatically. From eq. \metrica\ we get
$$
ds^2= A(\Psi_0)\big( 1-({g(q)\p_t\Psi_0\over A^2(\Psi_0)})^2\big) dt^2
-A^{-1}(\Psi_0)dq^2-2{g(q)\over A^2(\Psi_0 )}\p_t\Psi_0 dqdt \ \ .
\eqn\deese
$$

Eqs. \eqge \ or \effe \ can be interpreted as
the propagation of the  scalar
field $\ge $ in a nontrivial geometry. The source term $J_{\rm eff}$
in the right hand side of eqs. \eqge, \effe \ vanishes asymptotically
because we demand $\Psi _0 $ to satisfy the free field equation \free .

Now we would like to show explicitly that for a large class
of incoming waves this geometry corresponds to a black hole.
The simplest case is that in which $\Psi_0 $ is a static solution.
This case is no less unphysical than the $D=2$ {\it static} black hole
solution, but it illustrates some points.
{}From $\p_q^2\Psi_0 = 0$ it follows
$\p_q \Psi_0 $=const.
This can also be obtained from the exact solution
$$
{g_s\over \sqrt\pi }\p_x\Psi = \p_x\zeta _M -\p_x\zeta _0 \equiv
{1\over \pi} [(x^2-2M-1)^{1/2}-(x^2-1)^{1/2}]\cong
-{M\over \pi |x|} \ \ ,
\eqn\static
$$
or $\p_q \Psi  \cong -{M\over g}\ $. Thus we choose
$$
\p_q \Psi  _0 \equiv -{M\over g}\  \ \ .
\eqn\simple
$$
Inserting eq. \simple \ into eqs. \eqge , \deese \ we obtain

$$(A_0^{-1}\p_t^2 -A_0\p_q^2)\ge =J_0 +O(\ge ^2)\ ,\ \ J_0=O(M^2e^{-2q})
\ \ , \eqn\black $$
$$ds^2=A_0dt^2-A_0^{-1}dq^2 \  \ ,\ \ \
A_0\equiv 1-M{1\over \sinh^2(q)} \sim 1-4Me^{-2q} \ \ .
\eqn\hole
$$
This has the same form as the expression for the
Witten black hole. More exactly, after
a change of coordinates $\cosh(q)=\sqrt{1+M} \cosh(r) $
we obtain
$$ds^2=-dr^2 + \beta ^2(r)dt^2 \  ,\ \ \
\beta ^2(r)=(M+1) (M\coth^2(r)+1)^{-1} \ ,
\eqn\verl
$$
which is the `exact' metric found in ref. [\DVV ] (an exact solution to
the tree-level sigma-model $\beta $-function equations, see also
ref. [\ark ]). \foot {The fact that the exact DVV metric appeared
should actually be regarded as a fortunate `improvement' of the present
approximation. For example, a change of $\p_q\Psi_0 $ from the
constant value $\p_q\Psi_0= -{M\over g}$ to
$\p_q\Psi_0= -{M\over g} +O(e^{-2q}) $
would modify the metric by $O(e^{-4q})$. A generic feature of all
these DVV type metrics is the singularity at $q=0$.}

 In ref. [\wittbh ] the parameter $M$ was identified
with the ADM mass, which is usually assumed to be positive.
This geometry has a horizon at $\sinh^2(q_H)=M$ and a singularity at
$q=0$ or $x=1$, i.e. the position of the Polchinski wall (of course, the
present `linear' approximation breaks down much before getting
to the wall, so the singularity may just be an illusion for distant
observers).

Considering this particular form for $\Psi_0 $ is equivalent to
the expansion $\zeta=\zeta_M +{g_s\over\pi }\ge $ (see eqs.\fluc ,
\expa ). If we now introduce a coordinate $q_M$ as $x=
\sqrt{1+2M}\cosh(q_M) $
then we would obtain eq.\seq \ with $\ge $ instead of $\Psi $,
i.e. an equation of the form $(\p_{q_M}^2 - \p_t^2 )\ge =O(\ge ^2)$.
This of course agrees with eq.\black \ after changing $q\to q_M $,
which takes the metric $G_{\mu\nu}^{\rm eff}$ to the conformal gauge.
In this gauge the black hole geometry is not manifest, but the coordinate
$q_M$ is not geodesically complete since it does not cover the region
$\sqrt {1+2M}>x>1 $. In particular it does not include the horizon at
$x=\sqrt{M+1} $.

As a second example we consider a high pulse $\Psi_0=\Psi _{+}(x+t)$
coming from $x= \infty $ which extends above the line $p=|x|$.
The eigenvalues above the line will be on trajectories which carry them
over the barrier to $x<0$. For a very high pulse the reflected
part can be ignored. Consider for example the case in which
$\Psi _{+} $ is of the form, $\p_q\Psi _+(x+t)
= -E e^{-(x+t)^2}\ ,\ \ E<0$.
Then $A_0(\Psi_0) =1-{gE\over x^2-1 } e^{-(x+t)^2}$.
This geometry has  naked singularities at $x=\pm 1$.
The pulse may be interpreted as a wormhole connecting
the two asymptotically flat sides of the barrier.

Our third example is a low energy density pulse (by
`low energy density pulse' we mean a pulse which does not represent a large
deviation from the static solution, where some degenerate behaviours
can occur, leading to multivalued functions $p_\pm $ [\pol ]).
The exact tree-level $S$-matrix for these
pulses in the bosonic formalism was found in [\pol ] (for discussions
in the fermionic formulation see e.g. ref. [\GKLEB ]).
In the terms proportional to $\ge $
we can make use of the Polchinski exact solutions replacing
$\Psi _0$ by $\Psi $ since this does not affect the equation to
linear order in $\ge $. Replacing $\Psi_0$ by $\Psi $ in eqs. \metrica ,
 \deese \ and using eqs. \sol , \fluc \  we obtain
$$
G_{\rm eff}^{\mu\nu}=\left(\matrix{{2\sinh (q)\over p_+-p_-}&
{p_++p_-\over p_+-p_-} \cr
{p_++p_-\over p_+-p_-} &
{2p_+p_-\over (p_+-p_-)\sinh (q)} \cr}\right)
\eqn\metx
$$
or
$$
ds^2=-{2p_+p_-\over (p_+-p_-)\sqrt{x^2-1}} dt^2 - {2\sqrt{x^2-1}\over p_+-p_-}
dq^2 +2 {p_++p_-\over p_+ -p_-} dqdt
\eqn\mets
$$
Far away, $p_\pm=\pm x $, the $dqdt$ term vanishes and the metric
asymptotically approaches to the Minkowski metric
$\eta_{\mu\nu}$.
Eq. \mets\ provides a geometrical interpretation of the scattering
process for all $t $, and large $x$.
The metric \mets \
has a potential singularity at $x^2=1$
which may be absent in some specific cases. For example, if at
$x\cong 1 $ the momenta $p_\pm $ take its static values
$\sim \pm \sqrt{x^2-1}$
then the potential singularity cancels out. But if at some time $t$
a pulse is moving near the wall, $p_\pm (x\cong 1, t)$ will take
values very different
from the static case and  the metric can have a singularity at $x=1$,
presumably a curvature singularity (as viewed by a distant observer).
Thus the picture is the following: whenever a pulse is sent in,
an observer at large $x=x_0$ will see a time-dependent geometry
$G_{\rm eff}^{\mu\nu}(x\cong x_0,t)$ given by eq. \metx\ ,
which is curved in the regions where the pulse differs from the
static solution, and nearly flat elsewhere.
When the pulse is close to the wall, a singularity may develop
and the observer at $x_0$ may measure a large-distance black hole geometry.

To be more specific, let us consider the case of a `step' pulse
coming from $x=\infty $ and travelling anti-clockwise along
the Fermi surface in momentum space.
 When the pulse passes by, $p_+$ switches
from the static value $\sqrt{x^2-1} $ to $\sqrt{x^2-2M-1} \ ,\ \ M>0$.
After the step pulse has reached the wall, $p_+=\sqrt{x^2-2M-1} $
for all $x$. Then $p_-$ starts switching from its static
value $-\sqrt{x^2-1}$ to $-\sqrt{x^2-2M-1} $ as the step pulse
travels from $x=1$ to $x=\infty $ along the lower branch
of the Fermi surface. The time-depending
metric $G_{\rm eff}^{\mu\nu}(x,t)$ is
equal to $\eta_{\mu\nu} $ at the points $x$ which have
 not yet been reached by the pulse. It is
$$ds^2=A_1 dt^2- A_1^{-1} dq^2 -{M\over \sinh^2(q)}dqdt\ ,\ \ \
A_1\equiv 1-{M\over 2}{1\over \sinh^2(q)} \ \ ,
\eqn\paso
$$
at the $x$ which have been only reached by the ingoing step pulse
(we have dropped subleading terms in $e^{-2q}$, the exact
form can be read from eq. \mets ), and
$ds^2= {p_+\over \sinh(q)}dt^2-{\sinh(q)\over p_+} dq^2\cong
A_0 dt^2- A_0^{-1} dq^2 $ at the $x$ which have also
been reached by the outgoing step pulse
($A_0$ has been defined in eq.\hole ). This is the
metric everywhere for $t\to\infty $; it has a horizon at $p_+=0$,
i.e. $x^2_{\rm H}=2M+1$, and a singularity at $x=1$.

Now one is led to some speculation. Comparing with
$D=2$ continuum string theory black holes, we see that
in this scenario and in the large-distance approximation
$-\p_q\Psi_0$ plays the role of the integral
of an energy-momentum tensor (cf. e.g. eq. \constr).
If this energy-momentum tensor is positive-definite then this
integral is a monotonically nondecreasing quantity, which is
the basic reason why classical black holes can only increase
in size.
One could be tempted to demanding positivity on some of the
derivatives of $-\p_q\Psi _0$ to garantee that only positive
`energy density' is entering into the system.
This {\it ad hoc} restriction of the incoming pulses
would always
lead -in the classical theory- to metrics of the DVV type \hole \
 as final state.
In the case of the step pulse considered above $-\p_q\Psi _0$
is in fact monotonically
nondecreasing, but, in particular, in the case of localized pulses it is not.
After a localized pulse
is reflected and gets off from the wall, $p_\pm $
takes again values near the static solution and any seeming
black hole geometry evanesces.

When all higher powers of $\ge $ are incorporated
the exact scattering amplitude of low-energy pulses
is unitary and reveals no singularity or anomalous behaviour.
Perhaps this is a clue
of a secret reconciliation between black hole
physics and quantum mechanics.

\bigskip\bigskip
\noindent $\underline {\rm Acknowledgements}$: I wish to
thank W. Fischler, L. Susskind and A. Tseytlin
for useful discussions.

\refout
\end